%% file: main.tex
\documentclass[10pt,twocolumn,letterpaper]{article}
\usepackage{authblk}

\usepackage{cvpr}      
\usepackage{multirow}
\usepackage[safe]{tipa}
 

\definecolor{cvprblue}{rgb}{0.21,0.49,0.74}
\usepackage[pagebackref,breaklinks,colorlinks,allcolors=cvprblue]{hyperref}
\def\paperID{5715} 
\def\confName{CVPR}
\def\confYear{2025}

\title{StyleSpeaker: Audio-Enhanced Fine-Grained Style Modeling for Speech-Driven 3D Facial Animation}

\author[1]{An Yang}
\author[2]{Chenyu Liu}
\author[2]{Pengcheng Xia}
\author[1]{Jun Du\thanks{Corresponding author.}}
\affil[1]{NERC-SLIP, University of Science and Technology of China}
\affil[2]{IFLYTEK Research}

\begin{document}
\maketitle
\input{sec/0_abstract}    
\input{sec/1_introduction}
\input{sec/2_related_works}

\input{sec/3_method}
\input{sec/4_experiments}

\input{sec/5_conclusion}

{
    \small
    \bibliographystyle{ieeenat_fullname}
    \bibliography{main}
}


\end{document}

%% file: sec/0_abstract.tex
\begin{abstract}
\indent
Speech-driven 3D facial animation is challenging due to the diversity in speaking styles and the limited availability of 3D audio-visual data. Speech predominantly dictates the coarse motion trends of the lip region, while specific styles determine the details of lip motion and the overall facial expressions. 
Prior works lack fine-grained learning in style modeling and do not adequately consider style biases across varying speech conditions, which reduce the accuracy of style modeling and hamper the adaptation capability to unseen speakers.
To address this, we propose a novel framework, StyleSpeaker, which explicitly extracts speaking styles based on speaker characteristics while accounting for style biases caused by different speeches.
Specifically, we utilize a style encoder to capture speakers' styles from facial motions and enhance them according to motion preferences elicited by varying speech conditions.
The enhanced styles are then integrated into the coarse motion features via a style infusion module, which employs a set of style primitives to learn fine-grained style representation. Throughout training, we maintain this set of style primitives to comprehensively model the entire style space.
Hence, StyleSpeaker possesses robust style modeling capability for seen speakers and can rapidly adapt to unseen speakers without fine-tuning.
Additionally, we design a trend loss and a local contrastive loss to improve the synchronization between synthesized motions and speeches. 
Extensive qualitative and quantitative experiments on three public datasets demonstrate that our method outperforms existing state-of-the-art approaches.
\end{abstract}

%% file: sec/1_introduction.tex
\section{Introduction}
\label{sec:intro}
\begin{figure}[t]
\centering
\includegraphics[width=\linewidth]{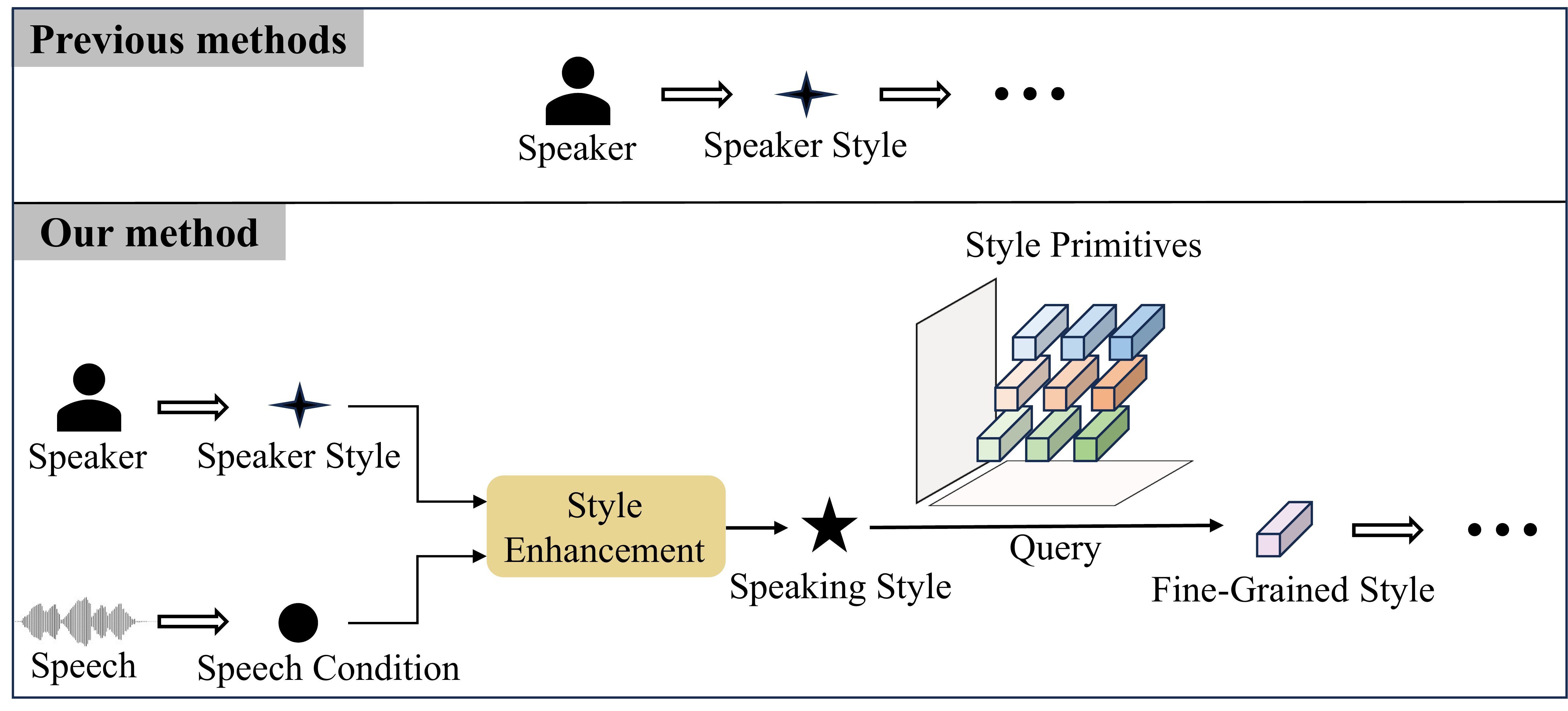}
\caption{Illustrations of the style modeling process of our proposed StyleSpeaker. Unlike previous methods, StyleSpeaker not only models speaking styles based on both speaker characteristics and specific speech, but also queries fine-grained representations based on style primitives.}
\label{style1}
\end{figure}


Speech-driven 3D facial animation has become an increasingly important research area due to its applications in virtual reality, film or game production, and biomimetic technology.
In this task, it is crucial to ensure not only the accuracy of lip motions but also the naturalness of the overall animation and the consistency of the speaking style characteristics.
The error in speech-driven 3D facial animation arises from three sources: viseme information provided by the speech, speaking style, and randomness. We focus only on the first two. The viseme information determines the coarse motion trends of the lip region, such as opening and closing, while the speaking style determines detailed lip motions and overall facial expressions.
The speaking style can be decomposed into two components: first, the speaker's inherent style, referred to as ``speaker style'' in the following sections, which stems from the speaker's habitual patterns. The speaker style dictates audio-independent motion habits, such as the motion patterns in the upper-face and the symmetry of lip motions during speech. Second, the specific speech condition, which further indicates motion preferences and influences the speaking style to some extent.
For example, speech intensity affects the maximum extent of mouth opening and the variance of changes in lip region. 
To summarize, the speaking style is primarily determined by the speaker identity but is influenced by the driven speech.

Accurately extracting the speaking style is crucial for synthesizing precise and characteristic-preserving facial animation. Previous works tend to overlook fine-grained style modeling and audio-induced style biases. Early methods~\cite{karras2017audio,richard2021audio} can only synthesize facial animation in a single style. Later works~\cite{cudeiro2019capture,richard2021meshtalk,fan2022faceformer,xing2023codetalker} embed one-hot encoding for individuals in the training set via an embedding layer and optimize this layer during training to implicitly capture speaker styles. However, the resulting styles become entangled with speech information, and cannot adapt to unseen speakers. Based on the prior methods, Imitator~\cite{thambiraja2023imitator} adapts to unseen speakers by fine-tuning some layers of the generalized model, which incurs additional overhead.
Recent methods~\cite{wu2023speech,fu2024mimic} begin to explicitly extract styles from facial motion sequences but ignore the impact of speech conditions and generalization learning in style space, which results in less accurate style extraction and weakens the adaptability to the styles of unseen speakers.

To address these issues, we propose a novel method, StyleSpeaker, which differs from previous methods by incorporating audio conditions into style considerations and using style primitives for fine-grained learning. We illustrate in Figure~\ref{style1} the improvements in the style modeling process of our method compared to previous methods.
Specifically, we employ a speaker style extractor to capture speaker styles.
Concurrently, we utilize an audio encoder to extract high-level audio features and learn audio conditions from them via an audio condition extractor. We then enhance the speaker styles with the audio conditions to produce speaking styles, aiming to integrate motion preference information embedded in the audio conditions into speaking styles.
Subsequently, we use a style infusion module that employs a set of style primitives to derive the fine-grained style representation for the speaking styles.
In practice, we deconstruct the input speaking styles into foundational representation using style primitives, which are simultaneously optimized throughout training.
Through iterative deconstruction and learning of encountered styles, these style primitives capture foundational style information to construct a comprehensive style space. For unseen speakers in the training set, we extract their speaking styles, map them within the style space for fine-grained representation, and acquire more precise style information. This approach significantly enhances our model's adaptability to diverse speaking styles.
In addition, we design two constraints: trend loss and local contrastive loss.
The trend loss imposes constraints on the higher-order differences in the facial motion sequences. Derived from CLIP~\cite{radford2021learning}, the local contrastive loss accounts for the recurrence of many syllables and facial motions within a speech segment. To avoid mismatches, we limit the computation of the contrastive loss to a local range. These two constraints significantly improve the accuracy of the synthesized animation.
We also design a new metric, Fourier Frequency Error, which effectively evaluates style consistency.
The main contributions of our work are as follows:
\begin{itemize}
\item We are the first to enhance speaking styles by incorporating audio conditions. Our fine-grained learning of the style space endows the model with strong style modeling capabilities, allowing rapid adaptation to unseen speakers without fine-tuning.

\item We introduce two novel constraint functions, trend loss and local contrastive loss, to further improve the accuracy and synchronization of the synthesized motions.

\item Extensive qualitative and quantitative experiments on three public datasets demonstrate that our method outperforms existing state-of-the-art approaches for both seen and unseen speakers.
\end{itemize}

%% file: sec/2_related_works.tex
\section{Related Works}
\label{sec:RelatedWorks}
\subsection{Speech-Driven 3D Facial Animation}
Initially, speech-driven 3D facial animation is synthesized using rule-based methods. 
The dominance functions~\cite{massaro200112} are employed to map speech to parameters controlling facial motions. Some methods~\cite{mattheyses2015audiovisual,edwards2016jali} model facial muscle motions from biological and anatomical perspectives and establish mappings with speech. 

With the advent of 4D face datasets, various learning-based methods have emerged~\cite{hwang2022audio}. Some methods~\cite{karras2017audio,richard2021audio} focus on driving a specific speaker, while more methods are dedicated to driving different speakers. VOCA~\cite{cudeiro2019capture} primarily generates lower face motions. MeshTalk~\cite{richard2021meshtalk} uses a categorical latent space to disentangle audio-correlated and audio-uncorrelated information. FaceFormer~\cite{fan2022faceformer} first employs the transformer architecture.
CodeTalker~\cite{xing2023codetalker} utilizes a discrete codebook approach to decouple motion space. FaceDiffuser~\cite{stan2023facediffuser}, DiffSpeaker~\cite{ma2024diffspeaker} and DiffSHEG~\cite{chen2024diffsheg} employ diffusion models. 
CorrTalk~\cite{chu2024corrtalk} divides faces into two regions based on audio correlation and uses two branches to generate motions separately. TalkingStyle~\cite{song2024talkingstyle} disentangles style codes from motion patterns and proposes a style-conditioned self-attention mechanism. EmoTalk~\cite{peng2023emotalk} utilizes a paired emotional content dataset to explicit control emotion styles. CSTalk~\cite{liang2024cstalk} proposes a parameter model for representing faces and learns the motion characteristics of specific emotions. 
All these methods use one-hot encoding to represent different speaker styles, which fails to generalize to unseen speakers.  
Subsequently, Imitator~\cite{thambiraja2023imitator} fine-tunes the generalized model for the adaptability to unseen speakers. ~\citet{wu2023speech} extracts styles from facial motions using the temporal convolutional architecture. Mimic~\cite{fu2024mimic} disentangles speaking styles and speech content using two separate latent spaces. DiffPoseTalk~\cite{sun2024diffposetalk} extracts styles for expressions and poses from parameterized facial sequences. ~\citet{yang2024probabilistic} propose a probabilistic model to preserve diversity in speech-driven tasks, while Probtalk3D~\cite{Probtalk3D_Wu_MIG24} employs a non-deterministic model with emotion control for similar purposes.
Recently, some works~\cite{eungi2024enhancing,zhao2024media2face,wu2024mmhead} utilize video or other modality information and pre-trained models to enhance the naturalness of synthesized 3D animation. 
The aforementioned methods overlook style variations under different speeches for a specific speaker and fail to learn style in the fine-grained manner, limiting the ability to model style and adapt to the new styles of unseen speakers.

\subsection{Stylized Talking Head Video Generation}
Various methods exist for stylized talking head video generation. ~\citet{ji2021audio} disentangle content and emotion information from audio and generate videos guided by the predicted landmarks. ~\citet{wang2020mead} and ~\citet{sinha2022emotion} use emotion labels as style codes but lack fine-grained individualized control. ~\citet{ji2022eamm} and ~\citet{liang2022expressive} use audio as content information and employ reference video frames as styles. StyleTalk++~\cite{wang2024styletalk++} and SyncTalk~\cite{peng2024synctalk} capture facial expression and head pose information from reference videos and extract them as style codes to control video generation. StyleSync~\cite{guan2023stylesync} encodes the style information into the $W^+$ space and generates target frames through StyleGan inversion\cite{abdal2019image2stylegan,abdal2020image2stylegan++,roich2022pivotal}.

%% file: sec/3_method.tex
\section{Method}
\subsection{Motivation}
\begin{figure}[t]
    \centering
    \includegraphics[width=\linewidth]{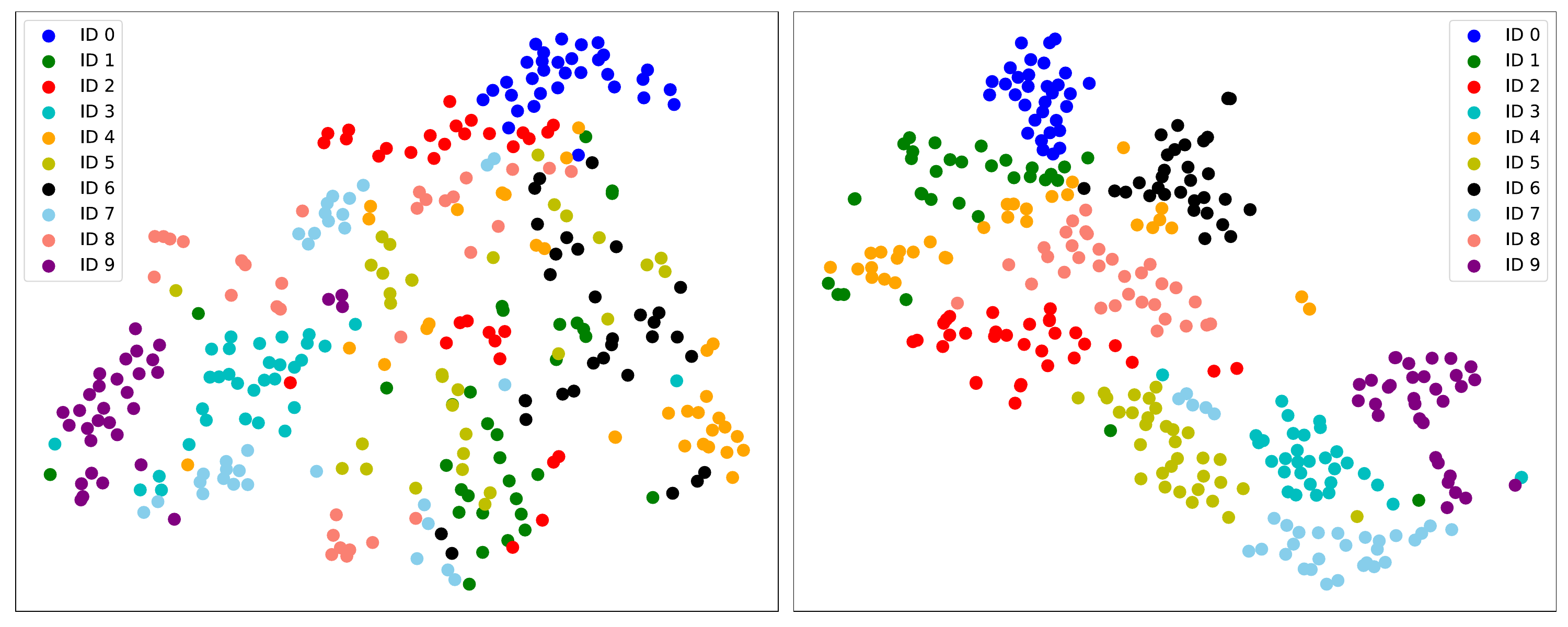}
    \begin{minipage}[t]{0.49\linewidth}
        \centering
        (a)
    \end{minipage}
    \hfill
    \begin{minipage}[t]{0.49\linewidth}
        \centering
        (b)
    \end{minipage}
    \caption{Visualization of motion characteristics under two feature extraction methods.}
    \label{stylespace1}
\end{figure}

To explore the distribution of speaking styles, we extract features from facial motion sequences of different speakers in different speeches. 
Inspired by FDD metric~\cite{xing2023codetalker}, we use upper-face dynamics deviation to reflect the speaking style information in facial motion sequences. Specifically, we calculate the standard deviation for each vertex motion sequence along the temporal dimension and project the results into 2D space for visualization using t-SNE~\cite{van2008visualizing}, as shown in Figure~\ref{stylespace1} (a). Additionally, we calculate the discrete Fourier transform and extract the first 20 principal frequency components for each vertex motion sequence in x, y, and z directions, and project the results into 2D space, as shown in Figure~\ref{stylespace1} (b).
Based on the two visualization results, we observe significant style differences among various speakers, while the same speaker exhibits different preferences under distinct speech conditions. This indicates that incorporating the style biases brought by audio can enhance the accuracy of style modeling. Moreover, variations across different speakers and speech conditions result in a multitude of speaking styles. To effectively handle the large variety of styles, we construct a style space using fundamental units to represent all styles, rather than extracting and directly applying styles as done in previous work. We aim for the fundamental units to learn and utilize all observed styles from the training phase as comprehensively as possible, becoming an expert capable of representing a broad spectrum of styles.
\subsection{Overview}
\begin{figure*}[t]
\centering
\includegraphics[width=1.0\textwidth]{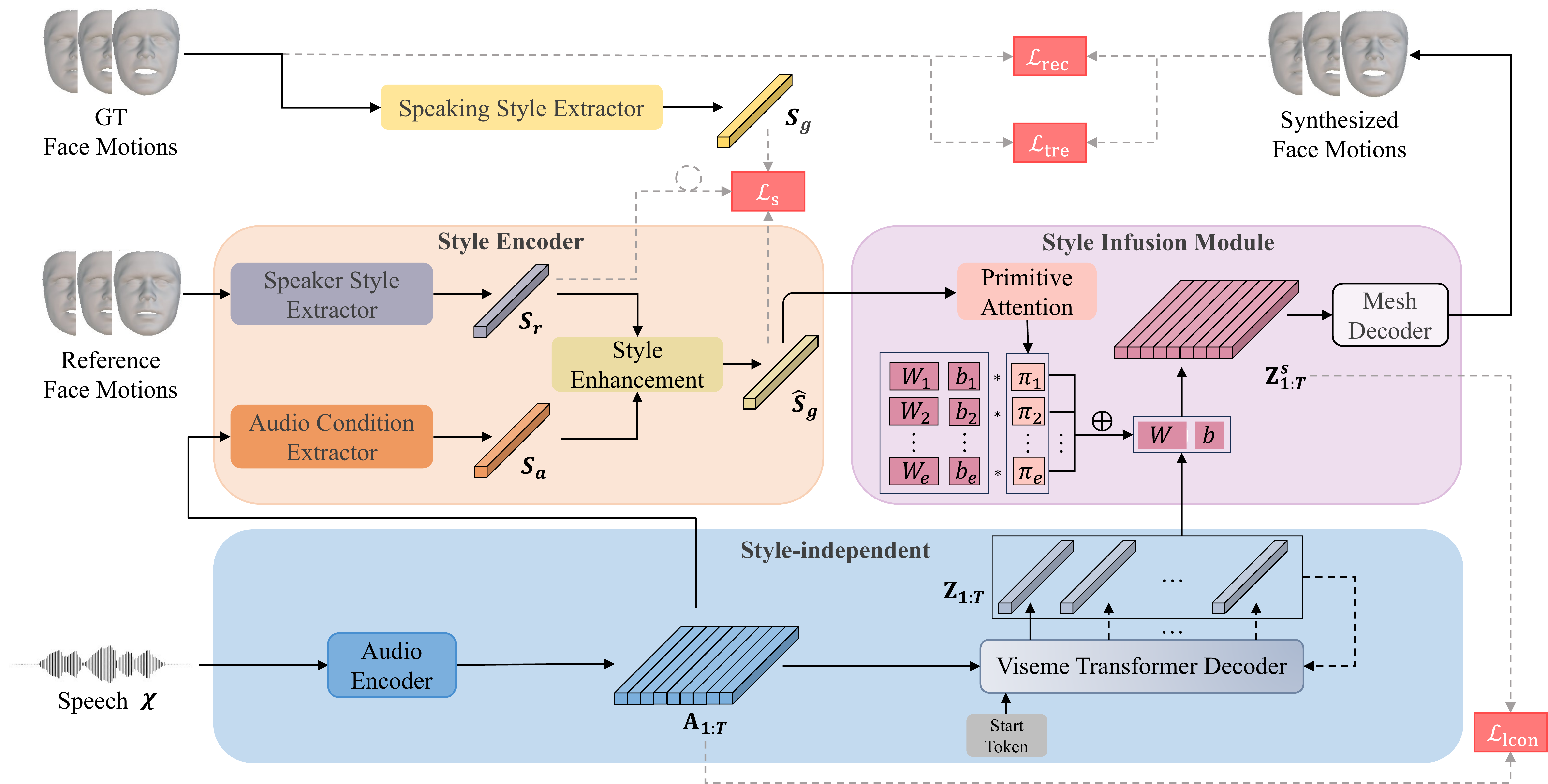} 
\caption{The pipeline of StyleSpeaker. Our framework separately extracts viseme features and style vectors before the final fusion. We use the audio encoder to extract audio features $\mathbf{A}_{1:T}$, which are then fed into the viseme transformer decoder to generate style-independent viseme features $\mathbf{Z}_{1:T}$ in an autoregressive manner. Concurrently, we extract the speaker style $S_{r}$ from reference
face motions and the audio condition vector $S_{a}$ from $\mathbf{A}_{1:T}$. We then use $S_{a}$ to enhance $S_{r}$, producing  the predicted speaking style $\hat{S}_{g}$. Finally, the style infusion module integrates $\hat{S}_{g}$ into $\mathbf{Z}_{1:T}$ using combined style primitives and synthesizes facial motions.}
\label{model1}
\end{figure*}

We focus on synthesizing 3D facial animation that is highly synchronized with speech and retains the characteristics of the speaking style.
To this end, we propose a novel model, StyleSpeaker, which comprises four modules: audio encoder, style encoder, viseme transformer decoder, and style infusion module, as shown in Figure~\ref{model1}. We use four constraint functions for training.
In the following, we will provide a detailed introduction to our model architecture, training strategy and objectives, and style adaptation.
\paragraph{Problem Formulation.}\vspace{-\baselineskip}
Let $\mathbf{M}_{1:T}=\{m_1,\dots ,m_T\}$ be a sequence of facial motions, where each frame $m_t \in \mathbb{R}^{N \times 3}$ denotes the displacement of $N$ vertices over a neutral-face mesh template $h \in \mathbb{R}^{N \times 3}$ in the xyz three directions at time $t$. Furthermore, let $\mathcal{X}$ be the speech segment. Our goal is to synthesize the facial motion sequence $\mathbf{M}_{1:T}$ based on the speech segment $\mathcal{X}$ and speaker information. Afterwards, we combine $\mathbf{M}_{1:T}$ with the template $h$ to obtain the target speech-driven 3D facial animation $\mathbf{O}_{1:T} = \{m_1 + h,\dots ,m_T + h\}$.

\subsection{Model Architecture}
\paragraph{Audio Encoder.}
The input audio $\mathcal{X}$ undergoes feature extraction before being fed into the subsequent network. 
We use a self-supervised pre-training model, WavLM~\cite{chen2022wavlm}, as our audio encoder, which consists of a convolutional feature encoder and a transformer encoder with gated relative position bias. We initialize our audio encoder with the pre-trained WavLM weights and freeze the convolutional feature encoder during training. 
To match the frame rate of the facial motions, we add a linear interpolation layer after the transformer encoder for resampling the audio features. We obtain the final audio features $\mathbf{A}_{1:T} = \{a_1,\dots,a_{T} \} \in \mathbb{R}^{d_{a} \times T}$.

\paragraph{Style Encoder.}\vspace{-\baselineskip}
To guide the synthesis of 3D facial animation, it is essential to capture the speaking style of the target speaker under the target speech, denoted as $S_g \in \mathbb{R}^{d_{s}}$. We decompose $S_g$ into two components: the speaker style vector $S_{r} \in \mathbb{R}^{d_{s}}$, which carries the speaker's typical speaking habits, and audio condition vector $S_{a} \in \mathbb{R}^{d_{s}}$, which carries the condition information of the target speech.
We extract $S_{r} $ from the facial motion sequence using an speaker style extractor, which consists of a temporal convolutional network (TCN) and feature extraction layers. The detailed structure is provided in Appendix A.
Given that the input facial motion sequences carry different speech conditions during training, we impose a consistency constraint on $S_{r} $ for the same speaker, ensuring that the speaker style extractor disregards these variations and captures the shared underlying information. 
We employ an audio condition extractor, structured similarly to the speaker style extractor, to learn $S_{a} $ from audio features $\mathbf{A}_{1:T}$. 
We then enhance $S_{r}$ with $S_{a}$ to derive the predicted speaking style $\hat{S}_g \in \mathbb{R}^{d_{s}}$.
The enhancement is as follows:
\begin{equation}
    S_\text{bias} =
    W_{s} \left[ \begin{array}{c}
    S_{r} \\
    S_{a}
    \end{array} \right] + b_{s} \label{style encoder},
\end{equation}
where $S_\text{bias}\in \mathbb{R}^{d_{s}}$ represents the style bias. $W_{s} \in \mathbb{R}^{d_{s} \times 2d_{s}}$ and $b_{s} \in \mathbb{R}^{d_{s}}$ are trainable parameters. We obtain $\hat{S}_{g}$ by:
\begin{equation}
\hat{S}_g = S_{r} + \alpha S_\text{bias} \label{con:S},
\end{equation}
where $\alpha$ is used to control the degree of the style bias. We set $\alpha = 0.1$ in our implementation. Additionally, we employ a speaking style extractor, with the same structure as the speaker style extractor, to learn the actual speaking style $S_g$ under the target speech as a form of supervision.

\paragraph{Viseme Transformer Decoder.}\vspace{-\baselineskip}
Our viseme transformer decoder is a multi-layer transformer decoder with biased causal self-attention to learn the dependencies of the current frame on past frames, and cross-modal attention to align the audio features $\mathbf{A}_{1:T}$ with the viseme features in the motion modality. We obtain the viseme features $\mathbf{Z}_{1:T} = \{z_1,\dots,z_{T} \} \in \mathbb{R}^{d_m \times T}$, which are independent of style and represent the coarse motions. Formally,
\begin{equation}
    \mathbf{Z}_{1:T} = D_v(\mathbf{A}_{1:T}).
\end{equation} 

\paragraph{Style Infusion Module.}\vspace{-0.5\baselineskip}
The style infusion module integrates the style information $\hat{S}_g$ into the viseme features $\mathbf{Z}_{1:T}$ and synthesizes the final facial motion sequence. 
Most previous methods directly input the extracted style as a reference into the subsequent network, whereas our approach incorporates fine-grained style learning and maintains a style primitive table during training, enhancing style adaptability.
Specifically, the module consists of a primitive attention layer, a style infusion layer, and a mesh decoder. 
We use the combination of $e$ pairs of base weights $(W_i, b_i)$ to project the style-agnostic $\mathbf{Z}_{1:T}$ to the style-infused motion features $\mathbf{Z}^s_{1:T}$, referring to these pairs of base weights as style primitives. We use these style primitives to decompose speaking styles into more basic representations, and aim to effectively represent diverse styles using the combinations of the style primitives. Specifically, $\hat{S}_g$ first undergoes the primitive attention layer to generate the attention weights for each primitive. These primitives are dynamically aggregated via the attention vector $\pi$, resulting in the style infusion layer weights $(W, b)$ corresponding to $\hat{S}_g$:
\begin{equation}
    W =\sum_{i=1}^e \pi_i W_i, b=\sum_{i=1}^e \pi_i b_i ,
\end{equation}
where $0 \leq \pi_i \leq 1$, $\sum_{i=1}^e \pi_i=1$. $\pi_i$ is the attention weight of i-th style primitive. $W \in \mathbb{R}^{d_m \times d_m}$, $b\in \mathbb{R}^{d_m}$. The style-infused motion features $\mathbf{Z}^s_{1:T} = \{z^s_1,\dots,z^s_{T} \} \in \mathbb{R}^{d_m \times T}$ are then generated by:
\begin{equation}
    z^s_{i} =  Wz_{i} + b.
\end{equation}
Finally, we obtain the predicted face motions $\hat{\mathbf{M}}_{1:T} = \{\hat{m}_1,\dots ,\hat{m}_T\}$, which are projected from $\mathbf{Z}^s_{1:T}$ via the mesh decoder.

\subsection{Training}
\paragraph{Training Strategy.}
For a target speaker and speech, we extract the speaker style $S_{r}$ from the reference facial motion sequence from other speeches of the same speaker, and learn the speech condition $S_{a}$ specific to the target speech.
These two components are then fused as a style reference $\hat{S}_g$ to synthesize the target motion sequence.
We impose consistency constraints on $S_{r}$ extracted from the same speaker to minimize its distance to the cluster center. Additionally, we extract the speaking style $S_g$ under the target speech condition from the ground-truth facial motion sequence to supervise the predicted speaking style $\hat{S}_g$, accelerating convergence. This training strategy effectively decouples the speaking style into the speaker style and audio-induced style biases.
\paragraph{Reconstruction Loss.} \vspace{-\baselineskip}
The reconstruction loss $\mathcal{L}_{\text{rec}}$ is:
\begin{equation}
    \mathcal{L}_{\text{rec}} = \sum_{t=1}^{T} \left \| \hat{m}_{t} - m_{t} \right \|^2_2.
\end{equation}

\paragraph{Style Loss.}\vspace{-0.5\baselineskip}
The style loss is defined as follows:
\begin{equation}
    \mathcal{L}_{\text{s}} = \big \| \hat{S}_{g} - S_{g} \big \|^2_2 +
    \big \| S_{r} - \mu (S_{r}) \big \|^2_2  ,
\end{equation}
where $\hat{S}_g$ is obtained from Equation (\ref{con:S}). Let $\mu (S_{r})$ represent the mean of all $S_{r}$ extracted from the reference facial sequences of the same speaker, with the one-hot encoding of the speaker in the training set as the initial value.
\paragraph{Trend Loss.}\vspace{-\baselineskip}
We impose constraints on the differences at various orders between facial motion sequences. Previous works use only the first-order difference as a constraint, but the facial motions between consecutive frames are minimal and do not effectively reflect the trend information. We believe that using differences at various time intervals can better constrain the facial motion trends. The trend loss is defined as follows:
\begin{equation}
    \mathcal{L}_{\text{tre}} = \frac{1}{R} \sum_{r=1}^{R} \sum_{t=1}^{T-r}  \left \|(\hat{m}_{t+r} - \hat{m}_{t})- (m_{t+r} - m_{t}) \right \|^2_2 ,
\end{equation}
where $R$ denotes the maximum order, we set $R=5$ in our implementation.
\paragraph{Local Contrastive Loss.}\vspace{-\baselineskip}
Inspired by CLIP~\cite{radford2021learning}, we design a local contrastive loss to align sequences with repetitive characteristics. In this work, we use this loss to align audio features $\mathbf{A}_{1:T}$ with motion features $\mathbf{Z}^s_{1:T}$, aiming to improve lip synchronization. The local contrastive loss is defined as follows:
\begin{equation}
    \mathcal{L}_{\text{lcon}} = \frac{1}{T} \sum_{t=1}^{T} \left ( \lambda\mathcal{L}_t^{(a \to m)} + (1-\lambda)\mathcal{L}_t^{(m \to a)}\right ) + \left \| W_l \right \|_1 ,
\end{equation}
where $\lambda = 0.5$. $\mathcal{L}_t^{(a \to m)}$ and $\mathcal{L}_t^{(m \to a)}$ represent audio-to-motion and motion-to-audio local contrastive loss:
\begin{align}
    \mathcal{L}_t^{(a \to m)}&=-\log \frac{\exp \left(\left\langle a_t, W_{l} z^s_t\right\rangle / \tau\right)}{\sum_{i=1}^T \exp \left(\left\langle a_t, W_{l}z^s_i\right\rangle / \tau\right) \cdot I^k_t(i)} ,
    \\\mathcal{L}_t^{(m \to a)}&=-\log \frac{\exp \left(\left\langle W_{l}z^s_t, a_t\right\rangle / \tau\right)}{\sum_{i=1}^T \exp \left(\left\langle W_{l}z^s_t, a_i \right\rangle / \tau\right) \cdot I^k_t(i)} ,
\end{align}
where $\langle\cdot\rangle$ denotes the cosine similarity. $\tau$ is the temperature parameter, which is fixed to 0.1 in our experiments. The matrix $W_l \in \mathbb{R}^{d_{a}\times d_m}$ is a set of learnable parameters that aligns $z^s_t$ from the motion space to the audio space. Since audio is strongly correlated only with the mouth region, we apply $l_1$ regularization constraint on $W_l$ to induce sparsity, encouraging the aligned features to focus on local regions. $I^k_t(i)$ is an indicator function that outputs 1 when $|i-t|\le k$, and 0 otherwise. This setup confines the computation of the contrastive loss within the range of $2k+1$ frames.

\paragraph{Training Objectives.}\vspace{-\baselineskip}
To train our model, we use $\mathcal{L}_{\text{total}}$ as our final loss function, defined as follows:
\begin{equation}
    \mathcal{L}_{\text{total}} = \lambda_{\text{rec}}\mathcal{L}_{\text{rec}} + \lambda_{\text{s}}\mathcal{L}_{\text{s}} +\lambda_{\text{tre}}\mathcal{L}_{\text{tre}} + \lambda_{\text{lcon}}\mathcal{L}_{\text{lcon}} .
\end{equation}
To ensure all loss terms remain at a similar scale, we set the weights as follows: $\lambda_{\text{rec}}=1.0$, $ \lambda_{\text{s}}=0.001$, $ \lambda_{\text{tre}}=1.0$, and $ \lambda_{\text{lcon}}=0.001$.

\subsection{Style Adaptation}
\label{sec:Style Adaptation}
Our model can rapidly adapt to the new styles of unseen speakers outside the training set. 
Given a video without audio of an unseen speaker, we obtain the 3D facial geometry sequence as a style reference sequence through a reconstruction method HRN~\cite{lei2023hierarchical}. We extract $S_{r}$ from the reference sequence using the trained speaker style extractor without fine-tuning the original model weights.


%% file: sec/4_experiments.tex
\section{Experiments}
\subsection{Datasets}
We use two widely adopted 4D datasets, BIWI~\cite{fanelli20103} and VOCASET~\cite{cudeiro2019capture}, along with the synthetic dataset 3D-MEAD. All three datasets provide paired spoken English audio and 3D facial geometry sequences.

\paragraph{BIWI Dataset.}\vspace{-\baselineskip}
BIWI is a corpus comprising affective speech and corresponding dense dynamic 3D face geometries. The dataset contains 40 sentences, each spoken by 14 subjects (8 females and 6 males) with an average duration of 4.67 seconds. 
The 3D face geometries are captured at 25 fps, each with 23370 vertices. Following~\citet{fan2022faceformer}, we select sentences with emotional context and partition the data as follows: a training set (BIWI-Train), a validation set (BIWI-Val), and two test sets (BIWI-Test-A and BIWI-Test-B). BIWI-Train, BIWI-Val, and BIWI-Test-A contain 192, 24, and 24 sentences, respectively, from the same 6 subjects. BIWI-Test-B contains 32 sentences from 8 unseen subjects.

\paragraph{VOCASET Dataset.}\vspace{-\baselineskip}
VOCASET contains 480 speech sentences and corresponding 3D facial geometry sequences from 12 subjects, with an average duration of about 4 seconds. The 3D facial geometries are captured at 60 fps, each with 5023 vertices. We follow the data split methodology of VOCA~\cite{cudeiro2019capture} to create a training set (VOCA-Train), a validation set (VOCA-Val), and a test set (VOCA-Test).

\paragraph{3D-MEAD Dataset.}\vspace{-\baselineskip}
3D-MEAD is synthesized through the 3D facial reconstruction method HRN~\cite{lei2023hierarchical} based on MEAD dataset~\cite{wang2020mead}. The 3D facial geometries are captured at 30 fps, each with 35709 vertices. More details are provided in Appendix B. 3D-MEAD comprises 1760 sequences from 44 speakers, which are divided into a training set (MEAD-Train), a validation set (MEAD-Val) and two test sets (MEAD-Test-A and MEAD-Test-B). MEAD-Train, MEAD-Val, and MEAD-Test-A contain 1152, 144, and 144 sequences, respectively, from the same 36 speakers. MEAD-Test-B contains 136 sequences from 8 unseen speakers. 

\subsection{Implementation Details}
Our framework is implemented by PyTorch. We train our model, StyleSpeaker, on a single NVIDIA A40 GPU for 50 epochs. We employ Adam optimizer for training, with an initial learning rate set to $0.0001$. After 40 epochs, the learning rate is decayed to 25$\%$ of the initial rate. 
We compare our method with FaceFormer~\cite{fan2022faceformer}, CodeTalker~\cite{xing2023codetalker}, FaceDiffuser~\cite{stan2023facediffuser}, CorrTalk~\cite{chu2024corrtalk}, Imitator~\cite{thambiraja2023imitator}, and Mimic~\cite{fu2024mimic}. 
More details of baselines and our implementation can be found in Appendix A.

\subsection{Quantitative Evaluation}
\paragraph{Metric.}
We quantitatively evaluate the synthesized motions for accuracy and style consistency. Currently, there is no widely accepted metric for style consistency evaluation. Previous studies use FDD~\cite{xing2023codetalker} for this purpose, but our observations in Figure~\ref{stylespace1} indicate that motion features extracted using the discrete Fourier transform provide better style distinction between different speakers. Therefore, we propose a new general metric, Fourier Frequency Error (FFE), for evaluating style consistency. Specifically, FFE is calculated by:
\begin{equation}
    \text{FFE}(\mathbf{M}_{1:T},\hat{\mathbf{M}}_{1:T}) = \frac{\sum_{c=1}^{N\times3} \big \|\mathcal{F}(\mathbf{M}_{1:T}^c) - \mathcal{F}(\hat{\mathbf{M}}_{1:T}^c)\big \|^2_2}{N\times3} ,
\end{equation}
where $\mathbf{M}_{1:T}^c \in \mathbb{R}^{T}$ denotes the sequence of a vertex motion component in x, y, or z direction. $\mathcal{F}$ represents the discrete Fourier transform, which extracts the first 20 principal frequency components.

We adopt four metrics: (1) Lip Vertex Error (LVE). It calculates the maximal $l_2$ error in the lip region between each frame of the synthesized motion sequences and the ground truth motion sequences, and averages this over all frames. (2) Face Vertex Error (FVE). It calculates the average $l_2$ error over the entire face between each frame of the synthesized motion sequences and the ground truth motion sequences, and averages this over all frames. (3) Face Dynamic Time Wrapping (FDTW). It evaluates synchronization by computing temporal sequence similarity using Dynamic Time Warping. (4) Fourier Frequency Error (FFE). 

\begin{table}[b]
\centering
\resizebox{1\columnwidth}{!}{
\begin{tabular}{lccccc}
\toprule
\multirow{2}{*}{Method} & LVE $\downarrow$ & FVE $\downarrow$ &  \multirow{2}{*}{FDTW $\downarrow$} & FFE $\downarrow$\\
& ($\times 10^{-4}$mm) & ($\times 10^{-5}$mm) & &($\times 10^{-2}$)  \\
\midrule
FaceFormer~\cite{fan2022faceformer}& 5.3077 & 8.7978 & 1.147 & 1.82 \\
CodeTalker~\cite{xing2023codetalker}& 4.7914 & 8.2758 & 1.156 & 1.63 \\
FaceDiffuser~\cite{stan2023facediffuser}& 4.2977 & 7.6533 & 1.097 & 1.46 \\
CorrTalk~\cite{chu2024corrtalk}& 4.0858 & 7.4356 & 1.090 & 1.37  \\
\textbf{Ours}& \textbf{3.8036} & \textbf{6.3635} & \textbf{0.995} & \textbf{1.26} \\
\bottomrule
\end{tabular}
}
\caption{Quantitative evaluations on BIWI-Test-A.}
\label{table1}
\end{table}

\paragraph{Comparison.} \vspace{-\baselineskip}
\label{sec:Comparison}
We first evaluate the synthesis capability of different models on BIWI-Test-A.
According to the results in Table~\ref{table1}, our model StyleSpeaker outperforms other models in terms of lip accuracy, overall facial motion accuracy and synchronization, and style consistency.
Moreover, to further compare the comprehensive synthesis capability on both seen and unseen speakers, we conduct experiments on 3D-MEAD, which encompasses a wider variety of styles. 
For MEAD-Test-B, we use a single motion sequence from the target speaker as a style reference.
According to the results in Table~\ref{table2}, our model outperforms Imitator and Mimic across all metrics on both MEAD-Test-A and MEAD-Test-B, particularly in LVE and FFE. This demonstrates that our model synthesizes more accurate lip motions and possesses stronger style adaptation capability.
\begin{table}[t]
\centering
\resizebox{1\columnwidth}{!}{
\begin{tabular}{lccccc}
\toprule
\multirow{2}{*}{Method} & LVE $\downarrow$ &  FVE $\downarrow$ &  \multirow{2}{*}{FDTW $\downarrow$} & FFE $\downarrow$\\
&($\times 10^{-3}$mm)&  ($\times 10^{-4}$mm) & &($\times 10^{-2}$)  \\
\midrule
Imitator~\cite{thambiraja2023imitator} & 1.2934 & 1.2748 & 1.204 & 2.29 \\
Mimic~\cite{fu2024mimic} & 1.2630 & 1.2194 & 1.221 & 2.29 \\
\textbf{Ours}& \textbf{0.9423} & \textbf{1.0391} & \textbf{1.167} & \textbf{1.87} \\
\midrule
Imitator   & 3.9451 & 4.5677 & 2.111 & 7.98  \\
$\text{Imitator}^{*}$   & 2.2391 & 2.2515 & 1.438 & 3.41  \\
Mimic   & 2.4400 & 2.3133 & 1.440& 3.42 \\
\textbf{Ours}  & \textbf{2.0236} & \textbf{2.1251} & \textbf{1.422} & \textbf{3.10} \\
\bottomrule
\end{tabular}
}
\caption{Quantitative evaluations on MEAD-Test-A (top 3 rows) and MEAD-Test-B (bottom 4 rows). $*$ denotes the fine-tuned Imitator.}
\label{table2}
\end{table}

\subsection{Qualitative Evaluation} 
\begin{figure*}[t]
\centering
\includegraphics[width=1.0\textwidth]{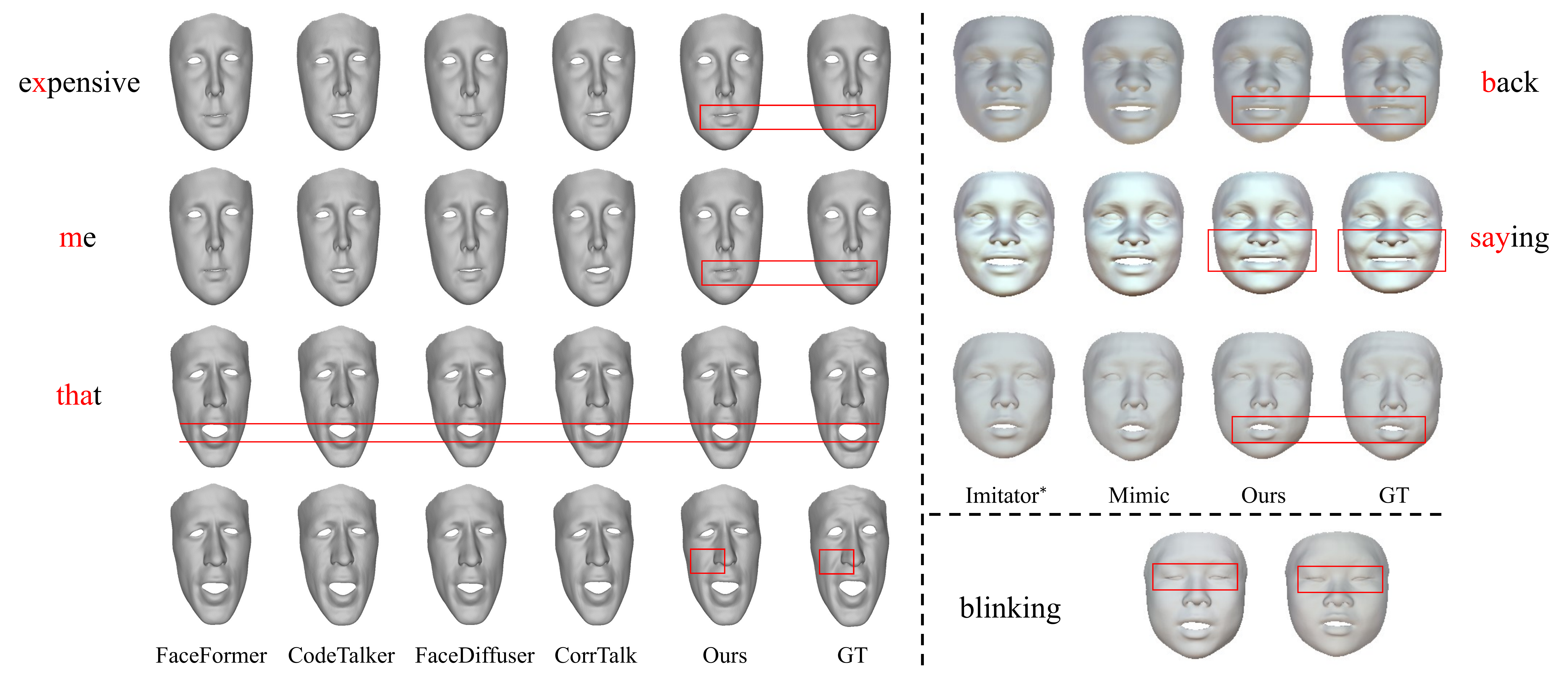} 
\caption{Visual comparisons with competing methods on BIWI-Test-A (left) and MEAD-Test-B (right).}
\label{vis}
\end{figure*}

\paragraph{Visual Comparison.}
We visually compare our method with competing methods in Figure~\ref{vis}. 
Our model synthesizes more accurate lip motions. For example, in the pronunciation of /s/ sound in ``expensive'', the synthesized lip motion from our model forms a narrow slit, closely matching the ground truth (GT). The lip closure during /m/ sound in ``me'' and the opening during /\textipa{D{a}}/ sound in ``that'' are also more closely aligned with the GT.
Additionally, our model captures more precise facial details for both seen and unseen speakers, such as preserving facial wrinkles and muscle contractions during speech, as well as the asymmetry of the lips when speaking. This highlights our model's stronger capabilities in style learning and adaptation.
Notably, even without applying specific constraints to the eye region, some of the synthesized motion sequences exhibit blinking actions, suggesting that periodic motions like blinking are effectively captured by our style modeling.
See the supplementary video for more visualization examples.



\paragraph{User Study.} \vspace{-\baselineskip}
To evaluate the performance of different methods perceptually, we conduct a user study on BIWI-Test-B, VOCASET-Test, and MEAD-Test-B, focusing on lip synchronization, realism, and style consistency. For BIWI-Test-B and VOCASET-Test, we obtain 30 videos for each method and create 150 A $\textit{vs.}$ B pairs (30 videos $\times$ 5 competing methods).
For MEAD-Test-B, we obtain 32 videos for each method and create 96 A $\textit{vs.}$ B pairs (with the GT videos for style evaluation).
We invite 30 participants with good vision and perception to complete the study.
Each pair is judged by at least 3 different participants, and 450, 450, 300 entries are collected for BIWI-Test-B, VOCASET-Test, and MEAD-Test-B. 
The percentage results indicate that our model achieves the best perceptual performance in terms of lip synchronization, realism, and style consistency, as shown in Table~\ref{table4}.


\begin{table}[t]
\centering
\resizebox{1\columnwidth}{!}{
\begin{tabular}{lcccc}
\toprule
\multirow{2}{*}{Ours vs. Competitors} & \multicolumn{2}{c}{BIWI-Test-B}  & \multicolumn{2}{c}{VOCA-Test}  \\
\cmidrule(r){2-3} \cmidrule(r){4-5}
& Lip Sync & Realism & Lip Sync & Realism  \\
\midrule
Ours vs. FaceFormer~\cite{fan2022faceformer}& 82.22 & 80.00 & 82.22 & 84.44 \\
Ours vs. CodeTalker~\cite{xing2023codetalker}& 76.67 & 75.56 & 80.00 & 72.22 \\
Ours vs. FaceDiffuser~\cite{stan2023facediffuser}& 66.67 & 63.33 & 68.89 & 67.78 \\
Ours vs. CorrTalk~\cite{chu2024corrtalk}& 65.56 & 55.56 & 58.89 & 63.33  \\
Ours vs. GT& 46.67 & 42.22 & 43.33 & 41.11 \\
\midrule
\multirow{2}{*}{Ours vs. Competitors} & \multicolumn{4}{c}{MEAD-Test-B}\\
\cmidrule(r){2-5}
& Lip Sync & Realism & \multicolumn{2}{c}{Style Consistency} \\
\midrule
Ours vs. $\text{Imitator}^{*}$~\cite{thambiraja2023imitator}& 71.00 & 83.00 & \multicolumn{2}{c}{68.00}  \\
Ours vs. Mimic~\cite{fu2024mimic}& 74.00 & 78.00 & \multicolumn{2}{c}{71.00}  \\
Ours vs. GT& 42.00 & 39.00 & \multicolumn{2}{c}{46.00} \\
\bottomrule
\end{tabular}
}
\caption{User study results on BIWI-Test-B, VOCA-Test, and MEAD-Test-B.}
\label{table4}
\end{table}
\subsection{Ablation Studies}
We conduct ablation experiments to assess the effectiveness of different modules and loss constraints, as shown in Table~\ref{table3}. 

\begin{table}[b]
\centering
\resizebox{1\columnwidth}{!}{
\begin{tabular}{lccccc}
\toprule
\multirow{2}{*}{Method} & LVE $\downarrow$ & FVE $\downarrow$ & \multirow{2}{*}{FDTW $\downarrow$} & FFE $\downarrow$\\
& ($\times 10^{-4}$mm) & ($\times 10^{-5}$mm) & & ($\times 10^{-2}$) \\
\midrule
Full      & \textbf{3.8036} & \textbf{6.3635} & \textbf{0.995} & \textbf{1.26}\\
w/o $\mathcal{L}_{\text{tre}}$       & 3.9428 & 6.5712 & 1.008 & 1.33 \\
with $\mathcal{L}_{\text{tre}}(R = 1)$       & 3.9236 & 6.5653 & 1.007 & 1.33 \\
w/o $\mathcal{L}_{\text{lcon}}$  & 3.9092 & 6.5564 & 1.006 & 1.31 \\
with contrastive loss  & 3.8801 & 6.5028 & 1.005 & 1.30 \\
w/o $S_a$ enhancement & 3.8943 & 6.4541 & 1.004 & 1.28 \\
w/o style primitives& 4.1017 & 6.7258 & 1.026 & 1.33 \\
\midrule
Full      & \textbf{2.0236} & \textbf{2.1251} & \textbf{1.422} & \textbf{3.10}\\
w/o style primitives& 2.1847 & 2.1973 & 1.429 & 3.30 \\
\bottomrule
\end{tabular}
}
\caption{Ablation study results on BIWI-Test-A (top 7 rows) and MEAD-Test-B (bottom 2 rows).}
\label{table3}
\end{table}

\paragraph{Constraints.} \vspace{-\baselineskip}
We observe a marked increase in all metrics after removing the trend loss, suggesting that motion trends over different time intervals encapsulate deep motion information and significantly enhance model performance. In contrast, applying the velocity loss from previous work yields only marginal improvements.
Similarly, removing the local contrastive loss results in notable metric increases, highlighting its role in aligning audio with synthesized motions. The comparisons with the contrastive loss demonstrate the necessity of our added local constraints.

\begin{figure}[t]
\centering
\includegraphics[width=\linewidth]{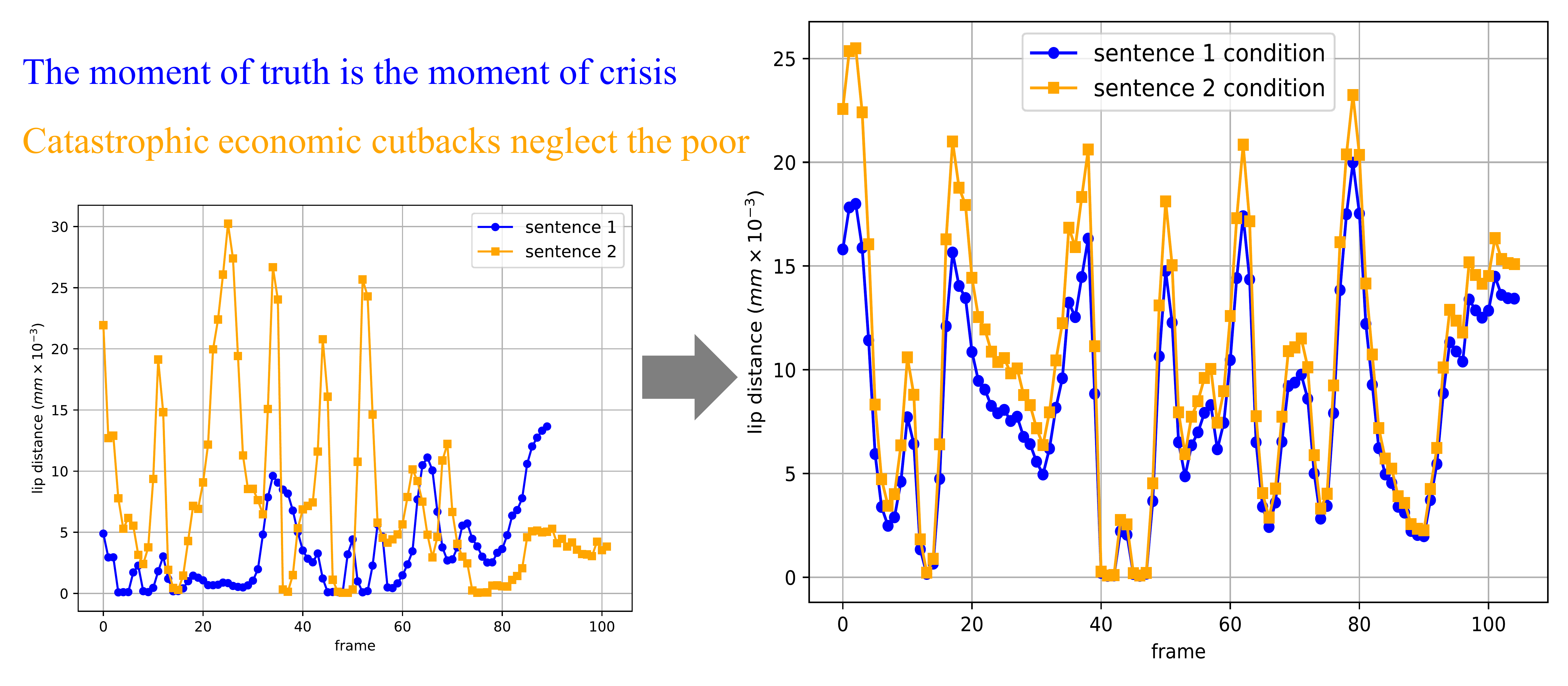} 
\caption{Comparisons of lip distance in synthesized animations under two audio conditions for the same speaker.}
\label{audioenhance}
\end{figure}

\paragraph{Audio Enhancement.} \vspace{-\baselineskip}
Audio enhancement further refines the extracted style by capturing motion preference details specific to the given speech condition based on the speaker style, making style information more accurate and enhancing the motion details in the synthesized facial animation. 
We select two speech samples from the same speaker in MEAD-Test-A, which correspond to large and slight lip articulations respectively, to enhance the speaker style. Based on these two speaking styles, we synthesize facial animations driven by the same speech and plot the lower-upper lip distances across frames for both styles in Figure~\ref{audioenhance}, from which we observe the subtle impact of the audio conditions on lip motions.

\begin{figure}[t]
\centering
\includegraphics[width=\linewidth]{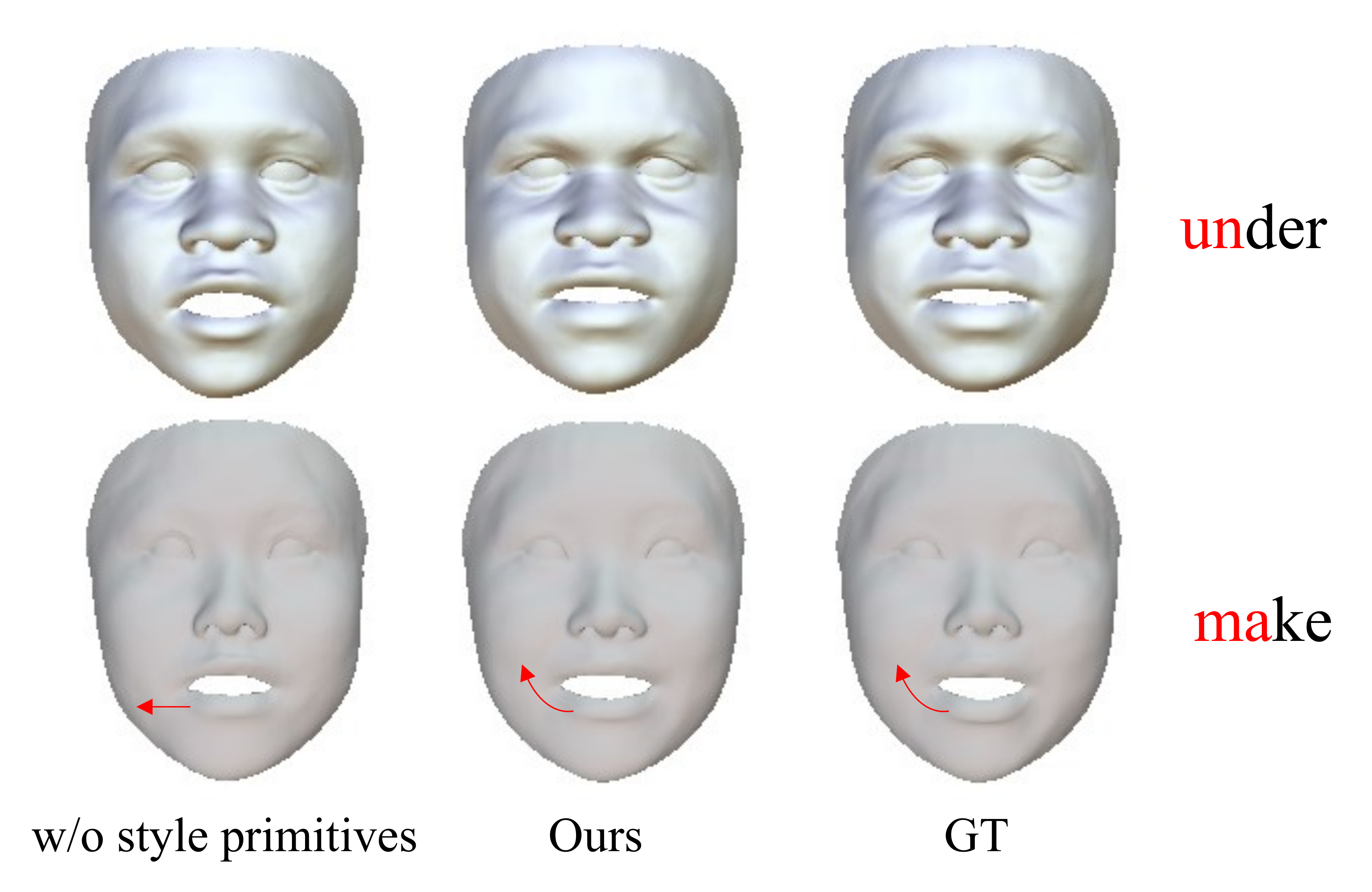} 
\caption{Visualization of the ablation study on the effect of the style primitives.}
\label{wosa}
\end{figure}

\paragraph{Style Primitives.} \vspace{-\baselineskip}
The style primitives decompose the speaking styles into more fundamental information, which significantly enhances the model's style learning and adaptation capabilities.
Results in Table~\ref{table3} demonstrate the crucial role of style primitives for both seen and unseen speakers. 
We present comparative examples of model outputs with and without style primitives in Figure~\ref{wosa}. 
Style primitives enhance large-scale style learning, such as the degree of mouth opening, while also refining small-scale details, such as the direction of mouth corners, resulting in more expressive and precise facial animation.

%% file: sec/5_conclusion.tex
\section{Conclusion}
In this paper, we propose StyleSpeaker, which achieves fine-grained style learning while accounting for audio-induced style biases, endowing our model with strong capabilities in style modeling and adaptation. 
Additionally, our proposed trend loss and local contrastive loss exhibit effectiveness in improving model performance.
Extensive experiments demonstrate that our model outperforms existing state-of-the-art methods in accuracy and style consistency on both seen and unseen speakers. However, the potential of our model for style modeling may not have been fully realized due to the limited variation in speech conditions within the dataset. 
In future work, we will focus on enhancing style modeling under varying conditions and emotions.
